\documentclass[twocolumn,showpacs,aps,prl,superscriptaddress]{revtex4}

\usepackage[dvips]{graphicx}
\usepackage{dcolumn}
\usepackage{epsfig}
\usepackage{amsmath}
\RequirePackage{xspace}

\input pubboard/babarsym
\RequirePackage{xspace}

\newcommand{\calP}{\ensuremath{{\cal P}}}

\newcommand{\pvec}{{\bf p}}

\newcommand{\acp}{\ensuremath{\calA_{ch}}}
\providecommand{\skz}{\mbox{$S$}}
\providecommand{\ckz}{\mbox{$C$}}
\def\deltaS{\ensuremath{{\rm \Delta}S}\xspace}
\newcommand{\calB}{\ensuremath{{\cal B}}}

\newcommand{\prodB}{\ensuremath{\prod\calB_i}}
\newcommand{\bfemsix}{\ensuremath{\calB(10^{-6})}}

\newcommand{\DE}{\ensuremath{\Delta E}}
\newcommand{\mb}{\ensuremath{m_{\rm ES}}}
\newcommand{\mres}{\ensuremath{m_{\rm res}}}
\newcommand{\xf}{\ensuremath{{\cal F}}}
\newcommand{\hel}{\ensuremath{{\cal H}}}

\providecommand{\dt}{\deltat}
\newcommand{\ttag}{\ensuremath{t_{\rm tag}}}
\newcommand{\bflav}{\ensuremath{B_{\rm flav}}}

\newcommand\etal{{\it et al.}}
\newcommand{\half}{\ensuremath{{1\over2}}}

\newcommand{\bfig}{\begin{figure}[htbpc!]}
\newcommand{\efig}{\end{figure}}
\newcommand\bef{\begin{figure}}
\newcommand\edf{\end{figure}}
\newcommand\dbline{\noalign{\vskip 0.10truecm\hrule}\noalign{\vskip 2pt}\noalign{\hrule\vskip 0.10truecm}}
\providecommand{\tbline}{\noalign{\vskip 0.05truecm\hrule\vskip0.05truecm}}

\newcommand\beq{\begin{equation}}
\newcommand\eeq{\end{equation}}
\newcommand\bear{\begin{array}}
\newcommand\enar{\end{array}}
\newcommand\beqa{\begin{eqnarray}}
\newcommand\eeqa{\end{eqnarray}}
\newcommand\ben{\begin{enumerate}}
\newcommand\een{\end{enumerate}}

\newcommand{\UfourS}{\ensuremath{\Upsilon(4S)}}

\newcommand{\omtoppp}{\ensuremath{{\omega\ra\pip\pim\piz}}}

\newcommand{\fomegapip}{\ensuremath{\omega\pi^+}}
\newcommand{\omegapip}{\ensuremath{\Bp\ra\fomegapip}}
\newcommand{\Bomegapip}{\ensuremath{\calB(\omegapip)}}
\newcommand{\romegapip}{\ensuremath{xx \pm xx\pm xx}}
\newcommand{\Romegapip}{\ensuremath{(\romegapip)\times 10^{-6}}}
\newcommand{\Aomegapip}{\ensuremath{0.xx\pm 0.yy \pm 0.zz}}

\newcommand{\somegapip}{\ensuremath{xx}}

\newcommand{\fomegaKp}{\ensuremath{\omega K^+}}
\newcommand{\omegaKp}{\ensuremath{\Bp\ra\fomegaKp}}
\newcommand{\BomegaKp}{\ensuremath{\calB(\omegaKp)}}
\newcommand{\romegaKp}{\ensuremath{xx\pm xx\pm xx}}
\newcommand{\RomegaKp}{\ensuremath{(\romegaKp)\times 10^{-6}}}
\newcommand{\AomegaKp}{\ensuremath{0.xx\pm 0.yy \pm 0.zz}}

\newcommand{\somegaKp}{\ensuremath{xx}}

\newcommand{\fomegaKz}{\ensuremath{\omega K^0}}
\newcommand{\fomegaKs}{\ensuremath{\omega\KS}}
\newcommand{\omegaKz}{\ensuremath{\Bz\ra\fomegaKz}}
\newcommand{\omegaKs}{\ensuremath{\Bz\ra\fomegaKs}}
\newcommand{\BomegaKz}{\ensuremath{\calB(\omegaKz)}}
\newcommand{\romegaKz}{\ensuremath{5.9^{+1.0}_{-0.9}\pm xx}}
\newcommand{\RomegaKz}{\ensuremath{(\romegaKz)\times 10^{-6}}}
\newcommand{\SomegaKz}{\ensuremath{0.xx^{+0.xx}_{-0.xx}\pm zz}}
\newcommand{\ComegaKz}{\ensuremath{-0.xx^{+0.xx}_{-0.xx}\pm zz}}
\newcommand{\somegaKz}{\ensuremath{xx}}

\renewcommand{\mres}{\ensuremath{m_{3\pi}}}
\providecommand{\sigdt}{\ensuremath{\sigma_{\deltat}}}
\renewcommand{\somegapip}{\ensuremath{10.8}}
\renewcommand{\somegaKp}{\ensuremath{13.0}}
\renewcommand{\somegaKz}{\ensuremath{8.6}}
\renewcommand{\romegapip}{\ensuremath{6.1\pm 0.7\pm 0.4}}
\renewcommand{\romegaKp}{\ensuremath{6.1\pm 0.6\pm 0.4}}
\renewcommand{\romegaKz}{\ensuremath{6.2\pm 1.0\pm 0.4}}
\renewcommand{\Aomegapip}{\ensuremath{-0.01\pm0.10\pm0.01}}
\renewcommand{\AomegaKp}{\ensuremath{0.05\pm0.09\pm0.01}}
\renewcommand{\SomegaKz}{\ensuremath{0.51^{+0.35}_{-0.39}\pm 0.02}}
\renewcommand{\ComegaKz}{\ensuremath{-0.55^{+0.28}_{-0.26}\pm 0.03}}

\newcommand{\BaBarYear}    {06}
\newcommand{\BaBarNumber}  {016}

\newcommand{\SLACPubNumber} {11777}

\begin{document}
\preprint{\babar-PUB-\BaBarYear/\BaBarNumber} 
\preprint{SLAC-PUB-\SLACPubNumber} 


\title{ \large \bf\boldmath Measurements of  $CP$-Violating Asymmetries
and Branching Fractions in $B$ Decays to $\omega K$ and $\omega\pi$}

%
\author{B.~Aubert}
\author{R.~Barate}
\author{M.~Bona}
\author{D.~Boutigny}
\author{F.~Couderc}
\author{Y.~Karyotakis}
\author{J.~P.~Lees}
\author{V.~Poireau}
\author{V.~Tisserand}
\author{A.~Zghiche}
\affiliation{Laboratoire de Physique des Particules, F-74941 Annecy-le-Vieux, France }
% ...
\collaboration{The \babar\ Collaboration}
\noaffiliation



\begin{abstract}
We present measurements of \CP-violating asymmetries and branching fractions 
for the decays \omegapip, \omegaKp, and \omegaKz.  The data 
sample corresponds to 232 million \BB\ pairs produced by \epem\ annihilation 
at the \UfourS\ resonance.  
For the decay \omegaKs, we measure the time-dependent \CP-violation parameters
$\skz = \SomegaKz$, and $\ckz = \ComegaKz$. 
We also measure the branching fractions, in units of $10^{-6}$,
$\Bomegapip=\romegapip$, $\BomegaKp=\romegaKp$, and $\BomegaKz=\romegaKz$,  
and charge asymmetries $\acp(\omegapip)=\Aomegapip$ and 
$\acp(\omegaKp)=\AomegaKp$.  
\end{abstract}

\pacs{13.25.Hw, 12.15.Hh, 11.30.Er}

\maketitle

Measurements of time-dependent \CP\ asymmetries in \Bz\ meson decays through
a Cabibbo-Kobayashi-Maskawa (CKM) favored $b \to c \bar{c} s$ amplitude
\cite{s2b,belles2b} have firmly established that \CP\ is not conserved in such
decays. The effect, arising from the interference between mixing and decay
involving the \CP-violating phase $\beta = \arg{(-V_{cd} V^*_{cb}/
V_{td} V^*_{tb})}$ of the CKM mixing matrix \cite{SM},
manifests itself as an asymmetry in the time evolution of the $\Bz\Bzb$ pair.   

Decays to the charmless final states $\phi\Kz$, $\Kp\Km\Kz$, $\etapr\Kz$,
$\piz\Kz$, $f_0(980)\Kz$, and $\omega\Kz$ are all $b\to \qqbar s$
processes dominated by a single penguin 
(loop) amplitude having the same weak phase $\beta$ \cite{Penguin}.  
CKM-suppressed amplitudes and multiple particles in the loop complicate 
the situation by introducing other weak phases whose contributions are not 
negligible; see Refs. \cite{Gross,BN} for early quantitative work in
addressing the size of these effects.  We define \deltaS\ as the difference 
between the time-dependent \CP-violating parameter $S$ (given in detail below) 
measured in these decays and $S=\stwob$ measured in charmonium $K^0$ decays. 
For the decay \omegaKz, these additional contributions are expected to
give $\deltaS\sim$0.1 \cite{beneke,CCS}, although this increase may be 
nullified when final-state interactions are included \cite{CCS}.
A value of \deltaS\ inconsistent with this expectation could be an indication of new physics \cite{lonsoni}.

We present an improved measurement of the time-dependent
\CP-violating asymmetry in the decay \omegaKz, previously reported by the
Belle Collaboration based on a sample of $\sim$30 events~\cite{BELLE}.  We also
measure branching fractions for the decays
\omegaKz, \omegapip, and \omegaKp\ (charge-conjugate decay modes are
implied throughout), and
for \omegapip, and \omegaKp, we measure the time-integrated charge asymmetry
$\acp = (\Gamma^--\Gamma^+)/(\Gamma^-+\Gamma^+)$, where $\Gamma^\pm$ is the 
width for these charged decay modes. In the Standard Model \acp\ is 
expected to be consistent with zero within our experimental uncertainty; 
a non-zero value would indicate direct \CP\ violation in this channel. 

The data were collected with the 
\babar\ detector~\cite{BABARNIM} at the PEP-II asymmetric $e^+e^-$ collider.
An integrated luminosity of 211~fb$^{-1}$, corresponding to
232 million \BB\ pairs, was recorded at the $\Upsilon (4S)$
resonance (center-of-mass energy $\sqrt{s}=10.58\ \gev$).
Charged particles are detected and their momenta measured by the
combination of a silicon vertex tracker (SVT), consisting of five layers
of double-sided detectors, and a 40-layer central drift chamber,
both operating in a 1.5 T axial magnetic field.  Charged-particle identification 
(PID) is provided by the energy loss in the tracking devices and by the
measured Cherenkov angle from an internally reflecting ring-imaging
Cherenkov detector (DIRC) covering the central region.
A $K/\pi$ separation of better than four standard deviations ($\sigma$)
is achieved for momenta below 3 \gevc, decreasing to 2.5$\sigma$ at the highest
momenta in the $B$ decay final states.
Photons and electrons are detected by a CsI(Tl) electromagnetic calorimeter.

From a $\Bz\Bzb$ pair produced in an \UfourS\ decay, we reconstruct one
of the $B$ mesons in the final state $f = \fomegaKs$, a \CP\ eigenstate
with eigenvalue $-1$.  For the time evolution measurement, we also
identify (tag) the flavor (\Bz\ or \Bzb) and reconstruct the
decay vertex of the other $B$.
The asymmetric beam configuration in the laboratory frame
provides a boost of $\beta\gamma = 0.56$ to the $\Upsilon(4S)$, which
allows the determination of the proper decay time difference $\dt \equiv
t_f-\ttag$ from the vertex separation of the two $B$ meson candidates.
Ignoring the \dt\ resolution (about 0.5 ps), the distribution of \dt\ is
\begin{eqnarray}
  F(\dt) &=& 
        \frac{e^{-\left|\deltat\right|/\tau}}{4\tau} [1 \mp\Delta w \pm
                                                   \label{eq:FCPdef}\\
   &&\hspace{-1em}(1-2w)\left( S\sin(\deltamd\deltat) -
C\cos(\deltamd\deltat)\right)].\nonumber
\end{eqnarray}
The upper (lower) sign denotes a decay accompanied by a \Bz (\Bzb) tag,
$\tau$ is the mean $\Bz$ lifetime, $\deltamd$ is the mixing
frequency, and the mistag parameters $w$ and
$\Delta w$ are the average and difference, respectively, of the probabilities
that a true $\Bz$\,($\Bzb$) meson is tagged as a $\Bzb$\,($\Bz$).
The parameter $C$ measures direct \CP\ violation.  If $C=0$, then
$S=\stwob+\deltaS$.

The flavor-tagging algorithm \cite{s2b} has seven mutually exclusive tagging categories
of differing purities (including one for untagged events that we
retain for yield determinations).  The measured analyzing power, defined as 
efficiency times $(1-2w)^2$ summed over all categories, is $( 30.5\pm 0.6)\%$,
as determined from a large sample 
of $B$-decays to fully reconstructed flavor eigenstates (\bflav).

We reconstruct a $B$ meson candidate by combining a \pip, \Kp\ or \KS\ with an 
\omtoppp.  We select $\KS\to\pi^+\pi^-$ decays by requiring the $\pi^+\pi^-$ 
invariant mass to be within 12 \mev\ of the nominal \Kz\ mass and by
requiring a flight length greater than three times its error.  We require the primary charged track to have a minimum of six Cherenkov photons in the DIRC.  We require the
$\pip\pim\piz$ invariant mass (\mres) to be between 735 and 825 \mev.  Distributions 
from the data and from Monte Carlo (MC) simulations \cite{geant}\ guide the 
choice of these selection criteria.  We retain regions adequate to
characterize the background as well as the signal for those quantities taken 
subsequently as observables for fitting.  We also use in the fit the angle 
$\theta_H$, defined, in the $\omega$ rest frame, as the angle of the
direction of the boost from the $B$ rest frame with respect to the normal to 
the $\omega$ decay plane.  The quantity $\hel\equiv|\cos{\theta_H}|$ is
approximately flat for background and distributed as $\cos^2{\theta_H}$ for signal.

A $B$ meson candidate is characterized kinematically by the energy-substituted mass
$\mes \equiv  \sqrt{(\half s + \pvec_0\cdot \pvec_B)^2/E_0^2 - \pvec_B^2}$ and the 
energy difference
$\DE \equiv E_B^*-\half\sqrt{s}$, where 
$(E_0,\pvec_0)$ and $(E_B,\pvec_B)$ are four-momenta of
the \UfourS\ and the $B$ candidate, respectively, and the asterisk 
denotes the \UfourS\ rest frame.  We require, assuming the \omegapip\ 
hypothesis, $|\DE|\le0.2$ GeV and $5.25\le\mes\le5.29\ \gev$.  

To reject the dominant background from continuum $\epem\ra\qqbar$ events 
($q=u,d,s,c$), we use
the angle $\theta_T$ between the thrust axis of the $B$ candidate and
that of
the rest of the tracks and neutral clusters in the event, calculated in
the  \UfourS\ rest frame.  The distribution of $\cos{\theta_T}$ is
sharply peaked near $\pm1$ for jet-like $q\bar q$
pairs and is nearly uniform for the isotropic $B$ decays; we require
$|\cos{\theta_T}|<0.9$ (0.8 for the charged $B$ decays).

From MC simulations of \BzBzb\ and \BpBm\ events, we find evidence 
for a small (0.5\%) \BB\ background contribution for the charged $B$ decays,
so we have added a \BB\ component to the fit described below
for those channels.

We use an unbinned, multivariate maximum-likelihood fit to extract
signal yields and \CP-violation parameters.  We use the discriminating 
variables \mes, \DE, \mres, \hel, and a Fisher 
discriminant \xf\ \cite{PRD}. The Fisher discriminant combines five 
variables: the polar angles with respect to the beam axis in the
\UfourS\ frame of the $B$ candidate momentum and of the $B$ thrust axis; 
the tagging category; and the zeroth and second 
angular moments of the energy flow, excluding the $B$ candidate, about the 
$B$ thrust axis \cite{PRD}.  We also use \dt\ for the \omegaKs\ decay,
while for the charged $B$ decays we use the PID variables $T_{\pi}$ and
$T_K$, defined as the number of standard deviations between the measured DIRC 
Cherenkov angle and that expected for pions and kaons, respectively. 

For the \omegaKs\ decay we define the probability density function
(PDF) for each event $i$, hypothesis $j$ (signal and \qqbar\ background),
and tagging category $c$ 
\begin{eqnarray}
\calP_{j,c}^i  &\equiv& \calP_j (\mes^i) \calP_j (\DE^i) \calP_j(\xf^i)
\calP_j (\mres^i) \\ \nonumber
&& \calP_j(\hel^i) \calP_j (\dt^i, \sigdt^i, c)\,, 
\end{eqnarray}
where $\sigdt^i$ is the error on \dt\ for event $i$.  
We write the extended likelihood function as
\begin{equation}
{\cal L} = \prod_{c} \exp{(-\sum_j Y_{j} f_{j,c})}
\prod_i^{N_c}\left[\sum_j Y_j f_{j,c} {\cal P}^i_{j,c}\right]\,,
\end{equation}
where $Y_j$ is the fit yield of events of species $j$, $f_{j,c}$ is the
fraction of events of species $j$ for each category $c$,
and $N_c$ is the number of events of category c in the sample. 
We fix $f_{{\rm sig},c}$ to $f_{\bflav,c}$, the values measured with the large
\bflav\ sample  \cite{s2b}.  The same likelihood function is used for
the charged decays except that the hypothesis $j$ also includes \BB\
background, the tagging category is not used and the
PDF is slightly different, involving flavor $k$ (primary \pip\ or \Kp):
\begin{eqnarray}
\calP^i_{jk} =  \calP_j (\mes^i) \calP_j (\DE^i_k,T^i_k)
\calP_j(\xf^i) \calP_j (\mres^i) \calP_j (\hel^i)\,.
\end{eqnarray}

The PDF $\calP_{\rm sig}(\dt,\, \sigdt, c)$, is the 
convolution of $F(\dt;\, c)$ (Eq.\ \ref{eq:FCPdef}) with the
signal resolution function (a sum of three Gaussians) determined from the
\bflav\ sample.
The other PDF forms are: the sum of two Gaussians for all signal shapes
except \hel, and the peaking component of the \mres\ background;
the sum of three Gaussians for 
$\calP_{\qqbar}(\dt; c)$; an asymmetric Gaussian with different
widths below and above the peak for $\calP_j(\xf)$ (a small ``tail"
Gaussian is added for $\calP_{\qqbar}(\xf)$); Chebyshev functions of
second to fourth order for \hel\ signal and the slowly-varying shapes of \DE, 
\mres, and \hel\ backgrounds; and, for $\calP_{\qqbar}(\mes)$, a
phase-space-motivated empirical function \cite{argus}, with a small Gaussian 
added for $\calP_{\BB}(\mes)$.

We determine the PDF parameters from simulation for the signal and \BB\
background components.  We study large control samples of $B\to D\pi$ decays 
of similar topology to verify the simulated
resolutions in \DE\ and \mes, adjusting the PDFs to account for any
differences found.  For the \qqbar\ background we use
(\mes,\,\DE) sideband data to obtain initial PDF-parameter values but ultimately 
leave them free to vary in the final fit.

We compute the branching fractions and charge asymmetry from fits performed
without \dt\ or flavor tagging.  The free fit parameters are 
the following: the signal and \qqbar\ background yields (the \BB\ yield, if 
present, is fixed); the three shape parameters of 
${\cal P}_{\qqbar}(\xf)$; the slope of ${\cal P}_{\qqbar}(\DE)$ and
${\cal P}_{\qqbar}(\mres)$; the fraction of the peaking component of
${\cal P}_{\qqbar}(\mres)$; $\xi$ \cite{argus}; and, for the charged $B$ decays,
the signal and background \acp.

\begin{table}[ht]
\caption{Fit sample size, signal yield, estimated yield bias (all in
events), estimated purity, detection efficiency,
daughter branching fraction product, statistical significance including
systematic errors, measured branching fraction, and
corrected signal charge asymmetry.}
\label{tab:results}
\begin{center}
\begin{tabular}{lccc}
\dbline
Quantity & \fomegapip & \fomegaKp & \fomegaKz  \\
\tbline
Events in fit&\multicolumn{2}{c}{44175}& 9145  \\
Signal yield&$274\pm28$&$266\pm24$ &$100\pm15$  \\
Yield bias        &  18  &  16  &   8  \\
Purity (\%)       &  34  &  46  &  46   \\
Eff. ($\epsilon$,\%) & 21.8 & 21.2 & 23.0  \\
$\prodB$             & 0.891 & 0.891 & 0.307  \\
$\epsilon\times\prodB$ (\%) & 18.2 & 17.7 &  6.4  \\
Significance ($\sigma$)     & \somegapip & \somegaKp & \somegaKz       \\
\bfemsix          & $6.1\pm0.7$&$6.1\pm0.6$& $6.2\pm1.0$     \\
Signal \acp       & $-0.01\pm0.10$~   & ~$0.05\pm0.09$&$-$  \\
\dbline
\end{tabular}
\end{center}
\vspace{-5mm}
\end{table}

Table \ref{tab:results} lists the quantities used to determine the branching 
fraction.  Equal production rates of \BpBm and \BzBzb pairs have been 
assumed.  Small yield biases are present in the fit, due primarily to
unmodeled correlations among the signal PDF parameters. In Table
\ref{tab:results} we include estimates of these biases, evaluated by
fitting simulated \qqbar\ experiments drawn from the PDF into which
we have embedded the expected number of signal and \BB\ background
events randomly extracted from the fully simulated MC samples.
The estimated purity in Table \ref{tab:results} is given by the
ratio of the signal yield to the effective background plus signal, the
latter being defined as the square of the error on the yield.

\begin{figure}[!htb]
 \includegraphics[angle=0,scale=0.4335]{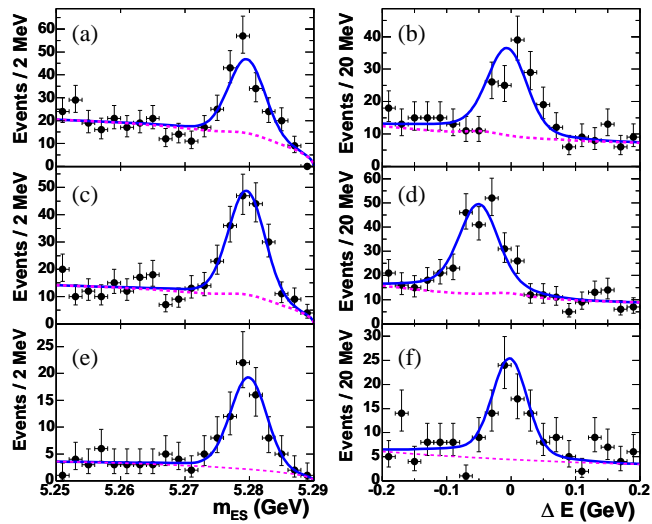}
\vspace{-.4cm}
 \caption{\label{fig:projMbDE}
The $B$ candidate \mb\ and \DE\ projections for \omegapip\ (a, b), \omegaKp\ 
(c, d), and \omegaKz\ (e, f) shown for a signal-enhanced subset of the data.  
Points with error bars represent the data, 
the solid line the fit function, and the dashed line the background components.
Note that the \fomegaKp\ signal in the \DE\ plot is displaced from zero
since \DE\ is defined for the \fomegapip\ hypothesis.}
\vspace{-.4cm}
\end{figure}

Fig.\ \ref{fig:projMbDE}\ shows projections onto \mb\ and \DE\ for
a subset of the data (including 45--65\% of signal events) 
for which the signal likelihood
(computed without the variable plotted) exceeds a 
threshold that optimizes the sensitivity.  

\begin{figure}[!tbp]
\begin{center}
\includegraphics[scale=0.4]{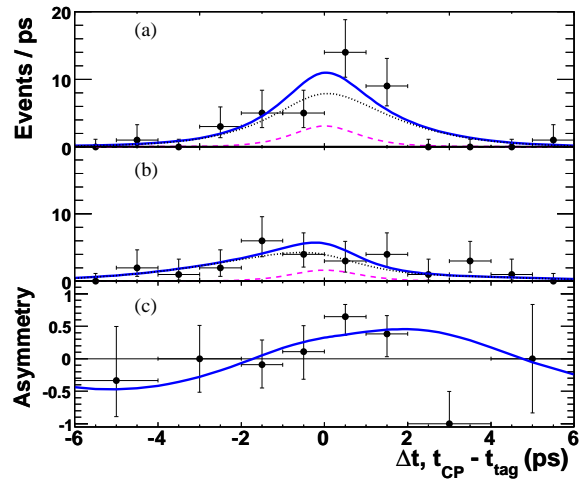}
\end{center}
\vspace{-.3cm}
\caption{Projections onto \deltat\ for \omegaKz.  Data (points with errors),
the fit function (solid line), background component (dashed line),
and signal component (dotted line), for events in which the tag meson is
(a) \Bz\ and (b) \Bzb, and (c) the
asymmetry $(N_{B^0}-N_{\Bbar^0})/(N_{B^0}+N_{\Bbar^0})$.}
\label{fig:dtproj}
\vspace{-0.4cm}
\end{figure}

For the time-dependent analysis, we require $|\dt|<20$ ps and $\sigdt<2.5$ ps.
The free parameters in the fit are the same as for the branching
fraction fit plus $S$, $C$, the fraction of background events in each
tagging category, and the six primary parameters describing the \dt\
background shape.  The parameters $\tau$ and 
$\deltamd$ are fixed to world-average values \cite{PDG2004}.
Here we find a slightly smaller yield of $95\pm14$ events and
$\skz=0.51^{+0.35}_{-0.39}$, $\ckz=-0.55^{+0.28}_{-0.26}$.
The errors have been scaled by $\sim$1.10 to account for a slight
underestimate of the fit errors predicted by our simulations when the
signal sample size is small.  Fig.~\ref{fig:dtproj} shows the $\Delta t$
projections and asymmetry of the time-dependent fit with
events selected as for Fig.~\ref{fig:projMbDE}.

The major systematic uncertainties affecting the branching fraction 
measurements include the reconstruction efficiency (0.8\%\ per charged track, 
1.5\%\ per photon, and 2.1\%\ per \KS) estimated from auxiliary studies.
We take one-half of the measured yield bias (3--4\%) as a systematic error.
The uncertainty due to the signal PDF description is estimated to be $\lsim$1\% 
in studies where the signal PDF parameters are varied within their estimated 
errors.  The uncertainty due to \BB\ background is also estimated to
be 1\% by variation of the fixed \BB\ yield by its estimated uncertainty.
The \acp\ bias is estimated to be $-0.005\pm0.010$ from studies 
of signal MC, control samples, and calculation of the asymmetry 
due to particles interacting in the detector.  We correct for this bias
and assign a systematic uncertainty of 0.01 for \acp\ for both
\omegapip\ and \omegaKp.

For the time-dependent measurements, we estimate systematic uncertainties 
in $S$ and $C$ due to \BB\ background and PDF shape variation (0.01 each), 
modeling of the signal \dt\ distribution (0.02), and interference between the 
CKM-suppressed $\bar{b}\to\bar{u} c\bar{d}$ amplitude and the favored 
$b\to c\bar{u}d$ amplitude for some tag-side $B$ decays \cite{dcsd}
(0.02 for $C$, negligible for $S$).  We also find that the uncertainty due
to SVT alignment and position and size of the beam spot are negligible.
The \bflav\ sample is used to determine the errors associated with the signal 
PDF parameters: \dt\ resolutions, tagging efficiencies, and mistag rates; 
published measurements \cite{PDG2004} are used for $\tau_B$ and \deltamd.  
Summing all systematic errors in quadrature, we obtain 0.02 for $S$ and
0.03 for $C$.

In conclusion, we have measured the branching fractions and time-integrated 
charge asymmetry for the decays \omegapip\ and \omegaKp\ and the
branching fraction for \omegaKz.  We find $\Bomegapip=\Romegapip$, 
$\BomegaKp=\RomegaKp$, $\BomegaKz=\RomegaKz$, $\acp(\omegapip)=\Aomegapip$,
and $\acp(\omegaKp)=\AomegaKp$, where the first errors are statistical
and the second systematic.  These results are substantially more precise than
earlier measurements \cite{BelleCLEO} and a significant improvement over our
previous measurements \cite{Previous}, which they supersede.
We also measure the time-dependent asymmetry parameters for the decay \omegaKz,
$\skz = \SomegaKz$ and $\ckz = \ComegaKz$, with a precision nearly a factor 
of two better than the previous Belle Collaboration results \cite{BELLE}.  If we fix
$C=0$, we find $S=0.60^{+0.42}_{-0.38}$.
This value of $\skz$ and the world-average value of
\stwob\ \cite{s2b,belles2b} yield a value of $\deltaS=0.12\pm0.40$, in
good agreement with the expected value near zero.

We are grateful for the excellent luminosity and machine conditions
provided by our \pep2\ colleagues, 
and for the substantial dedicated effort from
the computing organizations that support \babar.
The collaborating institutions wish to thank 
SLAC for its support and kind hospitality. 
This work is supported by
DOE
and NSF (USA),
NSERC (Canada),
IHEP (China),
CEA and
CNRS-IN2P3
(France),
BMBF and DFG
(Germany),
INFN (Italy),
FOM (The Netherlands),
NFR (Norway),
MIST (Russia),
MEC (Spain), and
PPARC (United Kingdom). 
Individuals have received support from the
Marie Curie EIF (European Union) and
the A.~P.~Sloan Foundation.

\end{document}